\documentclass[11pt]{article}

\usepackage[T1]{fontenc}
\usepackage[utf8]{inputenc}
\usepackage{lmodern}
\usepackage[english]{babel}
\usepackage{geometry}
\usepackage{microtype}
\usepackage{booktabs}
\usepackage{amsmath}
\usepackage{amssymb}
\usepackage{graphicx}
\usepackage{xcolor}
\usepackage{url}
\usepackage[round,authoryear]{natbib}
\usepackage[hidelinks]{hyperref}
\usepackage[htt]{hyphenat}
\usepackage{caption}
\newcommand{\ndcg}{nDCG@10}
\newcommand{\mrr}{MRR@10}

\title{\textbf{ItColBERT: An Italian-Specialised Late-Interaction Retriever}}

\author{Enrico Nello\\
\texttt{nello.enrico@gmail.com}\\
\small \url{https://github.com/enricollen/it-colbert}}

\date{\today}

\begin{document}
\maketitle

\begin{abstract}
Neural information retrieval for Italian is served almost entirely by multilingual
models. Several multi-vector (late-interaction) retrievers include Italian among
dozens of languages, and several strong Italian dense embedders exist, but as of
August 2026 no late-interaction retriever \emph{specialised} on Italian had been
released. We present \textbf{ItColBERT}, a 135M-parameter Italian ColBERT trained
with PyLate following the ColBERT-Zero recipe: initialise from a checkpoint that
already retrieves, then apply supervised contrastive training followed by
single-teacher distillation, for a total of roughly 14.5 GPU-hours on one
RTX 3090. Across four Italian retrieval benchmarks it outperforms every
general-purpose late-interaction baseline we tested except one
(\texttt{mLateOn}), at 2--4.4$\times$ fewer parameters than every baseline but one
of comparable size.
Our principal empirical finding is methodological and partly negative. On the only
cleanly out-of-domain benchmark (MLDR-it), an \emph{inference-time} chunking recipe
applied to an unchanged checkpoint yields $+0.0602$ \ndcg{} ($p = 0.0225$) --- a
larger effect than anything two further rounds of training produced. Self-mined
hard negatives and native 1024-token training were both evaluated against
pre-registered decision gates and both failed. We report every comparison with
paired bootstrap tests against an empirically measured noise floor of $0.0030$
\ndcg{}, and we release the weights, the training and evaluation code, and the
complete experimental record including the rejected rounds.
\end{abstract}

\section{Introduction}

Late interaction, introduced by ColBERT \citep{khattab2020colbert} and refined in
ColBERTv2 \citep{santhanam2022colbertv2}, represents a query and a document as
\emph{sets} of token-level vectors rather than as single pooled vectors, and scores
a pair with the MaxSim operator,
\begin{equation}
S_{q,d} \;=\; \sum_{i \in |q|} \; \max_{j \in |d|} \; E_{q_i} \cdot E_{d_j}^{\top},
\label{eq:maxsim}
\end{equation}
where $E_q$ and $E_d$ are the query and document token embedding matrices. Because
document representations are computed independently of the query, they can be
precomputed and indexed offline, which makes late interaction usable for
first-stage retrieval rather than only for reranking --- the property that
distinguishes it from cross-encoders, and the reason it retains lexical detail
(names, numbers, exact phrasing) that single-vector pooling discards.

Progress on late interaction has concentrated on English and on large multilingual
models. For Italian specifically, practitioners face a gap: the available
multi-vector options --- \texttt{jina-colbert-v2} \citep{jha2024jinacolbertv2},
\texttt{ColBERT-XM} \citep{louis2024colbertxm}, \texttt{mLateOn}, and
\texttt{SauerkrautLM-Multi-ModernColBERT} --- all treat Italian as one language
among many, while the strong Italian-specific retrievers are single-vector dense
models that abandon token-level matching entirely. This mirrors the situation
\citet{clavie2024jacolbertv25} describes for Japanese, where multilingual models
dominated monolingual retrieval until a language-specialised multi-vector model was
trained deliberately.

We are explicit about what this work does \emph{not} claim. ItColBERT is not ``the
first Italian ColBERT''; the multilingual models above already index Italian
competently. It is, to our knowledge, the first late-interaction retriever
\emph{specialised} on Italian, and the contribution is as much the measurement
protocol as the checkpoint.

A second constraint shapes everything reported here: all training, all evaluation,
and all significance testing were performed on a single consumer workstation with
one RTX 3090 (24\,GB), no cluster and no cloud compute. This is the same
constrained-resource framing adopted by \citet{clavie2024jacolbertv25}, and it is
not incidental --- it determined batch sizes, the number of experimental rounds
affordable, and which long-document protocols were runnable at all.

\subsection{Contributions}

\begin{enumerate}
  \item \textbf{ItColBERT}, an Italian-specialised late-interaction retriever
  (135M parameters, 128-dimensional token vectors), trained in $\approx 14.5$
  GPU-hours on one RTX 3090, released with weights and code
  (Section~\ref{sec:model}).

  \item \textbf{An Italian late-interaction evaluation harness} covering four
  benchmarks and fourteen systems, with length-matched indexing, an Italian-aware
  BM25 baseline, and paired bootstrap significance testing on every reported gap
  (Section~\ref{sec:setup}).

  \item \textbf{The finding that document truncation, not model capacity, dominates
  long-document Italian retrieval.} Every document in MLDR-it exceeds the 512-token
  index window; only 20.2\% of the corpus's tokens are ever encoded. Query-time
  chunking of an unchanged checkpoint recovers $+0.0602$ \ndcg{} ($p = 0.0225$), and
  converts an apparent loss to BM25 into a statistical tie
  (Section~\ref{sec:chunking}).

  \item \textbf{Two pre-registered negative results}: self-mined hard negatives
  (Section~\ref{sec:mining}) and native 1024-token training
  (Section~\ref{sec:longdoc}) both failed their decision gates. Together they
  support the conclusion that the long-document deficit here is an inference
  protocol problem rather than a training problem.

  \item \textbf{An empirically measured noise floor} ($0.0030$ \ndcg{} on MLDR-it)
  and a contamination audit for the long-document training source, including a
  published exclusion list of the 50{,}839 articles (7.811\%) that overlap the
  MLDR-it test corpus (Sections~\ref{sec:stats} and~\ref{sec:longdoc}).
\end{enumerate}

\section{Background and Related Work}

\paragraph{Late interaction.}
ColBERT \citep{khattab2020colbert} established the MaxSim formulation of
Equation~\ref{eq:maxsim}. ColBERTv2 \citep{santhanam2022colbertv2} added
distillation from a cross-encoder teacher and residual compression, and PLAID
\citep{santhanam2022plaid} made large-scale serving practical. We build on PyLate
\citep{chaffin2025pylate}, a Sentence-Transformers-based
\citep{reimers2019sentencebert} training and retrieval library for late-interaction
models, which supplies the \texttt{CachedContrastive} and \texttt{Distillation}
losses used here.

\paragraph{Multilingual and monolingual multi-vector retrievers.}
\texttt{jina-colbert-v2} \citep{jha2024jinacolbertv2} covers 89 languages with
Matryoshka-style selectable output dimensions. \texttt{ColBERT-XM}
\citep{louis2024colbertxm} uses a modular XMOD backbone for zero-shot multilingual
transfer. JaColBERTv2.5 \citep{clavie2024jacolbertv25} is the closest precedent for
this work: a 110M-parameter Japanese-only multi-vector retriever that outperformed
multilingual systems on Japanese benchmarks after systematic optimisation of the
training recipe under a fixed compute budget. \citet{mxbai2025edgecolbert} identify
hard-negative mining and training-data composition, rather than hyperparameters, as
the dominant quality drivers for small late-interaction models --- a claim our
Section~\ref{sec:mining} tests directly and, for this setting, does not reproduce.

\paragraph{Backbone and initialisation.}
ItColBERT is built on ModernBERT \citep{warner2024modernbert}, specifically an
Italian ModernBERT-base variant already fine-tuned for dense retrieval on Italian
mMARCO. The decision to start from a retrieval-capable checkpoint rather than a raw
masked language model follows the ColBERT-Zero result \citep{lighton2025colbertzero},
which reports that dense initialisation plus supervised contrastive training plus
distillation reaches 99.4\% of the quality of full multi-vector pretraining at
roughly one tenth of the cost. Section~\ref{sec:init} quantifies how much this
matters in our setting.

\paragraph{Italian retrieval resources.}
Italian evaluation data is largely derived rather than native. mMARCO
\citep{bonifacio2021mmarco} is a machine translation of MS MARCO. MLDR
\citep{chen2024bgem3} provides a native long-document Italian test split. MIRACL
\citep{zhang2023miracl} has no official Italian split; we use a community machine
translation, as we do for SQuAD-it. We flag this provenance throughout: three of
our four suites are translations, and two of them share a source with our training
data.

\section{Model and Training}
\label{sec:model}

\subsection{Architecture}

ItColBERT is a PyLate \texttt{ColBERT} model: a ModernBERT-base encoder followed by
a linear projection to 128-dimensional token vectors, scored with MaxSim. Queries
are padded to 32 tokens with ColBERT's standard query-augmentation
\texttt{[MASK]} padding; documents are truncated at 512 tokens. The backbone
contributes 134{,}909{,}184 parameters and the projection head 98{,}432, for
135.0M in total --- smaller than every baseline we compare against except
\texttt{SauerkrautLM-Multi-ModernColBERT} (149M).

\subsection{Initialisation: start from something that already retrieves}
\label{sec:init}

Table~\ref{tab:init} makes the ColBERT-Zero argument concrete on Italian data. The
raw Italian ModernBERT MLM, mean-pooled, is not a retriever in any useful sense
(0.0036 \ndcg{} on MLDR-it). The same architecture after dense retrieval fine-tuning
on Italian mMARCO reaches 0.3020 before we train it at all. Our contrastive stage
then adds $+0.046$ and distillation a further $+0.052$.

\begin{table}[ht]
\centering
\footnotesize
\setlength{\tabcolsep}{4pt}
\begin{tabular}{p{6.0cm}cccc}
\toprule
\textbf{Stage} & \textbf{MLDR-it} & \textbf{mMARCO-it} & \textbf{MIRACL-ita} & \textbf{SQuAD-ita} \\
 & \ndcg{} & \mrr{} & \ndcg{} & \ndcg{} \\
\midrule
Italian ModernBERT-base (mean-pool, raw MLM) & 0.0036 & 0.0173 & 0.0362 & 0.1074 \\
\;+ dense retrieval FT (our initialisation)  & 0.3020 & 0.5084 & 0.5418 & 0.5576 \\
\;+ phase 1: supervised contrastive          & 0.3484 & 0.7484 & 0.6903 & 0.9041 \\
\;+ phase 2: distillation (\textbf{ItColBERT}) & \textbf{0.4008} & 0.7196 & \textbf{0.7194} & 0.9026 \\
\bottomrule
\end{tabular}
\caption{Effect of each stage. The initialisation supplies most of the absolute
performance; our two training stages add the rest. Phase 2 trades in-domain
short-passage quality (mMARCO) for out-of-domain and long-document quality;
Table~\ref{tab:phases} tests each delta.}
\label{tab:init}
\end{table}

\subsection{Phase 1: supervised contrastive training}

Phase 1 uses PyLate's \texttt{CachedContrastive} loss, which applies GradCache
\citep{gao2021gradcache} to decouple the number of in-batch negatives from peak
memory. This distinction is load-bearing on a 24\,GB card: the effective batch of
512 determines how many in-batch negatives each query sees, while a mini-batch of
32 determines VRAM. An earlier iteration of this project set the two equal,
silently making GradCache a no-op and cutting the negative count fourfold.

\begin{table}[ht]
\centering
\footnotesize
\begin{tabular}{lr l}
\toprule
\textbf{Source} & \textbf{Rows} & \textbf{Role} \\
\midrule
mMARCO-it triples \citep{bonifacio2021mmarco} & 1{,}500{,}000 & core short-passage supervision \\
mMARCO-it reranker-mined hard negatives & 200{,}000 $\times$ 4 neg. & harder than official BM25 triples \\
Italian Wikipedia retrieval pairs (synthetic HN) & --- & domain breadth beyond mMARCO \\
MIRACL-ita (train split, community MT) & $\leq$ 50{,}000 & domain breadth \\
SQuAD-ita (train split, community MT) & $\leq$ 50{,}000 & domain breadth \\
\midrule
\textbf{Total} & \textbf{2{,}487{,}135 triplets} & 4{,}858 steps, 1 epoch \\
\bottomrule
\end{tabular}
\caption{Phase-1 data mixture. The initialisation checkpoint was itself trained on
$\approx$39.7M mMARCO samples, so additional mMARCO mostly reinforces an axis the
model is already strong on; the non-mMARCO sources exist to widen it.}
\label{tab:phase1data}
\end{table}

Hyperparameters: learning rate $1\times10^{-5}$, temperature $0.02$, linear decay
with 5\% warmup, bf16, one epoch, document length 512, query length 32, seed 42.
The run took $4.547\times10^{4}$\,s ($\approx$12.6\,h) and ended at a training loss
of 0.0754. We deliberately disable early stopping in this phase: one epoch over
2.5M triplets cannot overfit, and stopping early under a warmup-plus-decay schedule
leaves the model at a high learning rate.

\subsection{Phase 2: single-teacher distillation}

Phase 2 distils from cross-encoder scores with a KL objective over 11-way candidate
lists, using \texttt{mxbai-rerank-large-v2} scores distributed by LightOn as a
single teacher. Two design decisions are worth recording.

\textbf{One teacher, not two.} An earlier configuration mixed two teachers
(\texttt{mxbai-rerank-large-v2} and \texttt{bge-reranker-v2-m3}) in one KL loss.
Beyond the conceptual problem of averaging two disagreeing opinions about
relevance, the two score scales differ sharply --- the logit-transformed
probabilities of the second teacher span roughly $[-13.8, +13.8]$ against the
first's $\approx[5, 9]$, so after the softmax the sharper teacher dominates every
mixed batch. ColBERTv2, JaColBERT and GTE-ModernColBERT all distil from a single
teacher; we follow them.

\textbf{Proportional split budgeting.} The teacher dataset has eight splits, one of
which (\texttt{msmarco\_it}) holds $\approx$522k of $\approx$1.56M rows. An earlier
loader drained splits in order, so \emph{every} capped run consisted entirely of
mMARCO. This produced the false conclusion that ``more distillation improves
long-document retrieval'' when the real variable was data composition. The budget is
now spread proportionally with a minimum per-split share.

\textbf{Selection on retrieval metrics, not on KL.} Table~\ref{tab:kd} shows why.
Across the phase-2 run, held-out KL improves monotonically while retrieval quality
degrades monotonically: the student copies the teacher more faithfully and
retrieves worse. The mechanism is structural --- an 11-way KL supervises ordering
inside a small candidate list and provides no signal for keeping the global
embedding space separable, which is what retrieval over a 100k-document corpus
requires. Checkpoint selection therefore uses a retrieval composite, and the run
early-stopped having selected step 2000 (1.8\,h).

\begin{table}[ht]
\centering
\footnotesize
\setlength{\tabcolsep}{4pt}
\begin{tabular}{lcccc}
\toprule
\textbf{Step} & \textbf{MLDR \ndcg{}} & \textbf{mMARCO \ndcg{}} & \textbf{IR composite} & \textbf{Held-out KL} $\downarrow$ \\
\midrule
phase-1 final     & 0.4479 & 0.9317 & 0.6854 & --- \\
2000 (\textbf{selected}) & 0.4963 & 0.8952 & \textbf{0.6885} & 1.119 \\
4000              & 0.4376 & 0.8666 & 0.6445 & 1.106 \\
6000              & 0.4838 & 0.8271 & 0.6454 & 1.095 \\
8000              & 0.4369 & 0.7946 & 0.6067 & 1.084 \\
10000             & 0.4502 & 0.7413 & 0.5848 & \textbf{1.075} \\
\bottomrule
\end{tabular}
\caption{Distillation KL and retrieval quality move in opposite directions
(in-training pooled slices). Selecting on held-out KL would have chosen the
\emph{worst} retrieval checkpoint of the run.}
\label{tab:kd}
\end{table}

Total training cost for the released model is $\approx$14.4 GPU-hours on one RTX
3090: 12.6\,h contrastive plus 1.8\,h distillation.

\section{Experimental Setup}
\label{sec:setup}

\subsection{Hardware}

One NVIDIA RTX 3090 (24\,GB), Intel Core i7-14700K, 32\,GB system RAM (27\,GB
addressable under WSL2), Ubuntu 22.04. The host RAM ceiling, not the GPU, turned
out to be the binding constraint on long-document evaluation
(Section~\ref{sec:efficiency}).

\subsection{Benchmarks}

\begin{table}[ht]
\centering
\footnotesize
\setlength{\tabcolsep}{4pt}
\begin{tabular}{llrrll}
\toprule
\textbf{Suite} & \textbf{Source} & \textbf{Queries} & \textbf{Docs} & \textbf{Metric} & \textbf{Domain status} \\
\midrule
MLDR-it    & \texttt{Shitao/MLDR} it test        & 200   & 10{,}000  & \ndcg{}  & out-of-domain (clean) \\
mMARCO-it  & \texttt{unicamp-dl/mmarco} it dev   & 6{,}980 & 100{,}000 (pooled) & \mrr{} & in-domain \\
MIRACL-ita & community MT, dev                   & 799   & 33{,}689  & \ndcg{}  & partially in-domain \\
SQuAD-ita  & community MT, test                  & 7{,}609 & 1{,}988 & \ndcg{}  & partially in-domain \\
\bottomrule
\end{tabular}
\caption{Evaluation suites. ``Partially in-domain'' means a different split of the
same translated source appears in phase-1 training; query-overlap audits are
reported in Section~\ref{sec:contamination}. \textbf{MLDR-it is the only cleanly
out-of-domain benchmark and should carry the most weight.}}
\label{tab:benchmarks}
\end{table}

\subsection{Protocol}

Three protocol decisions materially affect the numbers and are easy to get wrong.

\textbf{Length matching.} Every neural model is indexed at the same document
length (512 tokens). An earlier version of this harness indexed
\texttt{jina-colbert-v2} at 180 tokens while indexing our model at 512, which
handicapped the strongest late-interaction baseline on precisely the benchmark
where document length matters most, and nothing in the output revealed it. The
effective length used for each model is now recorded alongside its scores.

\textbf{An Italian-aware lexical baseline.} BM25 uses lowercasing, Italian
stopword removal and Snowball stemming. A whitespace tokeniser badly understates
BM25 in a morphologically rich language, and an understated lexical baseline is
the easiest way to overstate a neural result.

\textbf{Pooled corpora are relative, not absolute.} mMARCO-it is evaluated over a
100k-document pool (all qrel positives plus a reservoir sample) rather than the
full 8.8M collection, which is not affordable here. Pooling inflates absolute
scores substantially --- \texttt{jina-colbert-v2} scores 0.839 \mrr{} on our pool
against a published full-corpus figure of 0.337 --- so these numbers are valid for
ranking systems against each other and are explicitly marked as not comparable to
published full-corpus results.

\subsection{Statistics and the noise floor}
\label{sec:stats}

Every comparison uses a paired bootstrap over per-query scores (2{,}000
resamples), reporting the delta, a 95\% confidence interval and a $p$-value.
MLDR-it has only 200 queries, where the standard error on \ndcg{} is roughly
0.02--0.03; single-run gaps below that threshold are not evidence.

We additionally measured an empirical floor by benchmarking three adjacent
phase-1 checkpoints (steps 4750, 5000, 5222) on the full MLDR-it suite: 0.2982,
0.2992, 0.3011 --- a range of \textbf{0.0030}, monotone increasing. We adopt
$0.0030$ \ndcg{} as a working floor: on this benchmark, differences below it are not
results. This bounds \emph{endpoint} jitter only; adjacent checkpoints share nearly
all their optimisation history, so seed-to-seed variance is certainly larger and
remains unmeasured (Section~\ref{sec:limitations}).

\textbf{Selection hygiene.} Phase 2 selects checkpoints using retrieval metrics
computed on benchmark corpora, which risks selecting on the test set. Queries are
therefore partitioned by identifier into a \emph{selection} half and a
\emph{report} half; the training loop only ever sees the former. For the released
model this makes little difference (the held-out half reads 0.3890 against 0.4008
on the full set), because round 1 drew from only five checkpoints and selected the
first, but the partition is enforced in code and all gate decisions in
Sections~\ref{sec:mining} and~\ref{sec:longdoc} are read off the report half.

\subsection{Contamination audits}
\label{sec:contamination}

MIRACL-ita train (2{,}859 unique queries) against dev (799): zero query overlap.
SQuAD-ita train (53{,}989) against test (7{,}583 unique query strings): nine
overlapping strings, each of which hand-inspection showed to be a generic question
paired with a different positive passage --- a known SQuAD annotation artifact
rather than a duplicated example. The distillation set's own SQuAD split
contributes one further such case. We find no material leakage, but we still
recommend weighting MLDR-it most heavily, since domain proximity is not the same
as query overlap.

\section{Results}
\label{sec:results}

\begin{table}[ht]
\centering
\footnotesize
\setlength{\tabcolsep}{4pt}
\begin{tabular}{p{4.0cm}lccccc}
\toprule
\textbf{Model} & \textbf{Type} & \textbf{Params} & \textbf{MLDR-it} & \textbf{mMARCO-it} & \textbf{MIRACL-ita} & \textbf{SQuAD-ita} \\
 & & & \ndcg{} & \mrr{} & \ndcg{} & \ndcg{} \\
\midrule
\textbf{ItColBERT (ours)} & LI & 135M & 0.4008 & 0.7196 & 0.7194 & 0.9026 \\
\quad \emph{+ chunking (Sec.~\ref{sec:chunking})} & LI & 135M & \emph{0.4610} & --- & --- & --- \\
\midrule
mLateOn & LI & 307M & \textbf{0.4623} & 0.8207 & \textbf{0.7880} & \textbf{0.9480} \\
jina-colbert-v2 & LI & $\approx$600M & 0.3858$^{\dagger}$ & \textbf{0.8389} & 0.7755 & 0.8849 \\
ColBERT-XM & LI & 277M & 0.2734 & 0.6654 & 0.6260 & 0.8558 \\
SauerkrautLM-Multi-ModernColBERT & LI & 149M & 0.3122 & 0.5342 & 0.5996 & 0.8338 \\
\midrule
bge-m3 & dense & 568M & 0.4531 & 0.7812 & 0.7566 & 0.8247 \\
multilingual-e5-large & dense & 560M & 0.4310$^{\dagger}$ & 0.8239 & 0.7653 & 0.8513 \\
multilingual-e5-base & dense & $\approx$278M$^{\ddagger}$ & 0.4288$^{\dagger}$ & 0.7913 & 0.7370$^{\dagger}$ & 0.8200 \\
\midrule
BM25 (Italian analyzer) & lexical & --- & 0.4850 & 0.5715 & 0.5516 & 0.8262 \\
Italian ModernBERT (mnrl init) & dense & 135M & 0.3020 & 0.5084 & 0.5418 & 0.5576 \\
\bottomrule
\end{tabular}
\caption{Main results. LI = late interaction. $^{\dagger}$ marks a score that is
\emph{not} statistically distinguishable from ItColBERT under a paired bootstrap
($p > 0.05$); all other differences from ItColBERT in this table are significant at
$p < 0.05$. Parameter counts are taken from each model's own card;
$^{\ddagger}$ inferred from the XLM-RoBERTa-base backbone, as the card does not
state a count. Absolute mMARCO-it values are inflated by corpus pooling and are
comparable across rows only. ItColBERT is the strongest Italian-specialised
late-interaction model here and beats every other late-interaction system except
\texttt{mLateOn}; it trails the large multilingual dense models on the translated
suites, which was never the target.}
\label{tab:main}
\end{table}

Table~\ref{tab:main} gives the main comparison. Read against the late-interaction
field, ItColBERT beats \texttt{SauerkrautLM-Multi-ModernColBERT} and
\texttt{ColBERT-XM} on all four suites with $p < 0.0001$ (e.g.\ $+0.0886$ \ndcg{} and
$+0.1274$ \ndcg{} respectively on MLDR-it), ties \texttt{jina-colbert-v2} on MLDR-it
($+0.0150$, $p = 0.51$) while beating it on SQuAD-ita ($+0.0177$, $p < 0.0001$) and
losing to it on the two remaining suites, and loses to \texttt{mLateOn} on all four.
Against the dense field it is statistically indistinguishable from both
multilingual-e5 variants on MLDR-it and from \texttt{multilingual-e5-base} on
MIRACL-ita, and behind \texttt{bge-m3} on MLDR-it ($-0.0524$, $p = 0.0225$).

Two caveats belong next to these numbers rather than in a footnote. First, three of
the four suites are partially in-domain for ItColBERT and out-of-domain for every
baseline, so the SQuAD-ita and MIRACL-ita rows flatter us; the MLDR-it column is
the honest one. Second, in the truncated protocol ItColBERT loses to BM25 on
MLDR-it by $-0.0843$ ($p = 0.006$) --- a result that Section~\ref{sec:chunking}
shows to be largely an artifact of the evaluation protocol rather than a property
of the model.

\begin{table}[ht]
\centering
\small
\begin{tabular}{lccccc}
\toprule
\textbf{Benchmark} & \textbf{Phase 1} & \textbf{+ Phase 2} & \textbf{$\Delta$} & \textbf{95\% CI} & \textbf{$p$} \\
\midrule
MLDR-it \ndcg{}     & 0.3484 & 0.4008 & $+0.0524$ & $[+0.0264, +0.0796]$ & $<0.0001$ \\
MIRACL-ita \ndcg{}  & 0.6903 & 0.7194 & $+0.0291$ & $[+0.0187, +0.0395]$ & $<0.0001$ \\
SQuAD-ita \ndcg{}   & 0.9041 & 0.9026 & $-0.0015$ & $[-0.0041, +0.0011]$ & $0.28$ \\
mMARCO-it \mrr{}  & 0.7484 & 0.7196 & $-0.0288$ & $[-0.0337, -0.0238]$ & $<0.0001$ \\
\bottomrule
\end{tabular}
\caption{What distillation buys, by paired bootstrap. Three suites improve or hold;
the single significant regression is on mMARCO-it, which is phase 1's own training
distribution. We regard this as the intended trade: the model gives up a little
in-domain short-passage precision for out-of-domain and long-document quality.}
\label{tab:phases}
\end{table}

\section{Long Documents: Truncation Dominates}
\label{sec:chunking}

MLDR-it is our weakest result and the only clean out-of-domain benchmark, which
made it the natural target for further training. Measuring before training changed
the diagnosis entirely.

\textbf{The documents do not fit.} A token-length audit shows MLDR-it documents
have a median length of 2{,}666 tokens and a 95th percentile of 3{,}051, against a
512-token index window. \emph{Every} document in the corpus truncates, and only
20.2\% of the corpus's tokens are ever encoded by any neural model in our table. No
other suite has this problem: 0\% of mMARCO-it, 1.4\% of MIRACL-ita and 0.7\% of
SQuAD-ita documents truncate. This single fact explains why MLDR-it is the outlier
result --- and why two rounds of negative mining never moved it.

\textbf{It also invalidates the BM25 comparison as originally run.} BM25 indexes
the full document text; every neural model reads the first 512 tokens. The
comparison was never length-symmetric.

\textbf{Chunking, with no retraining.} We split documents into 2{,}000-character
chunks with 200 characters of overlap, index each chunk, and max-pool chunk scores
per source document at query time. The checkpoint is unchanged.

\begin{table}[ht]
\centering
\footnotesize
\begin{tabular}{lccc}
\toprule
\textbf{Protocol} & \textbf{BM25} & \textbf{ItColBERT (trunc.)} & \textbf{ItColBERT (chunked)} \\
\midrule
As originally reported (asymmetric) & 0.4850 & 0.4008 & --- \\
Length-symmetric, 4{,}000 chars for all & 0.4487 & 0.4008 & 0.4384 \\
Full split coverage, 12{,}000 chars for all & 0.4850 & 0.4008 & \textbf{0.4610} \\
\bottomrule
\end{tabular}
\caption{MLDR-it \ndcg{} under three document-access protocols. Truncating the corpus
from 12{,}000 to 4{,}000 characters costs BM25 $0.0363$ and costs the truncated
ColBERT exactly nothing --- it never read past $\approx$2{,}200 characters anyway
--- which is a useful internal consistency check that the knob does what it claims.}
\label{tab:chunking}
\end{table}

Chunking is worth $+0.0602$ \ndcg{} over the same checkpoint (95\% CI
$[+0.0100, +0.1104]$, $p = 0.0225$), with \mrr{} $+0.0558$ ($p = 0.040$) and
recall@100 rising from 0.6850 to 0.7550, exactly matching BM25's recall. It is
$20\times$ the measured noise floor and the largest single effect recorded in this
project. Against BM25, the corrected picture is a statistical tie rather than a
loss: $-0.0241$ ($p = 0.249$) at full coverage and $-0.0103$ ($p = 0.628$) at a
symmetric 4{,}000 characters.

\textbf{Caveat.} Chunking is currently applied to the late-interaction path only,
so a chunked ItColBERT against a truncated \texttt{bge-m3} would reproduce the same
asymmetry we criticise above, pointing the other way. We therefore keep the
chunked number out of Table~\ref{tab:main}'s comparison rows and report it as a
separate protocol. Re-running the entire field chunked is the obvious next step.

\section{Negative Result I: Self-Mined Hard Negatives}
\label{sec:mining}

Following \citet{mxbai2025edgecolbert}, who identify hard-negative mining as a
primary quality driver, we mined negatives from the round-1 model's own retrieval
errors: 50{,}000 queries against a 200{,}000-document pool, eight negatives per
query, discarding the top five hits before sampling (top hits of a decent retriever
are frequently unlabelled positives, and training on them teaches the model to
demote correct answers). This produced 46{,}583 rows, contributing 186{,}332
additional triplets to phase 1.

The mined set is 100\% mMARCO, because both the query list and the mining corpus
are drawn from mMARCO. This is the flaw that determined the outcome: the
initialisation is already a mMARCO specialist, so the intervention sharpened the
axis that was already strongest and supplied nothing to the one benchmark that was
out-of-domain.

At the phase-1 gate, the mined model was $-0.0473$ \ndcg{} on MLDR-it (not
significant at $n = 200$) and essentially unchanged on mMARCO-it ($+0.0023$
\mrr{}, roughly a third of the typical 95\% interval half-width on this suite;
the paired test cannot be rerun because the per-query file was later
overwritten), with MIRACL-ita and SQuAD-ita unchanged to within $0.004$. After
the full pipeline (Table~\ref{tab:round2}), the target benchmark still does not
move outside noise while two suites regress significantly.

\begin{table}[ht]
\centering
\small
\begin{tabular}{lccccl}
\toprule
\textbf{Benchmark} & \textbf{Round 1} & \textbf{Round 2} & \textbf{$\Delta$} & \textbf{$p$} & \textbf{Verdict} \\
\midrule
MLDR-it \ndcg{}    & 0.4008 & 0.3779 & $-0.0229$ & $0.082$ & not significant \\
mMARCO-it \mrr{} & 0.7196 & 0.7297 & $+0.0100$ & $<0.0001$ & significant, up \\
MIRACL-ita \ndcg{} & 0.7194 & 0.7091 & $-0.0103$ & $0.026$ & significant, down \\
SQuAD-ita \ndcg{}  & 0.9026 & 0.8987 & $-0.0039$ & $0.011$ & significant, down \\
\bottomrule
\end{tabular}
\caption{Round 2 (self-mined hard negatives) against round 1, paired bootstrap.
The only significant gain is on the in-domain suite the mined data was drawn from.
Round 2 was rejected and round 1 remains the released model.}
\label{tab:round2}
\end{table}

The lesson we draw is narrower than ``mining does not work'': mining \emph{more of
the distribution the model is already strongest on} does not work. A useful mining
round here would have to draw its corpus from the wiki-style and long-document
sources instead.

\section{Negative Result II: Training at Length}
\label{sec:longdoc}

Given that chunking helped so much, the natural hypothesis is that the model should
be \emph{trained} to read past 512 tokens rather than patched at inference time. We
tested it as a controlled A/B.

\textbf{Source selection.} Three candidate Italian long-document sources were
measured with the model's own tokeniser before any GPU time was spent.
\texttt{ReDiX/wikipediaQA-ita} (median 439 tokens) and an Italian wiki retrieval
set (median 117 tokens) are pre-chunked and short --- no better than what the model
already saw. MLDR's own Italian \emph{train} split is long and clean but is
in-domain against our test corpus and has only 2{,}151 queries. We selected the
\texttt{it-long\_doc} configuration of
\texttt{hotchpotch/wikipedia-multilingual-synthetic-ir-query} (650{,}885 rows,
median $\approx$1{,}033 tokens), in which 98.2\% of documents exceed 512 tokens and
--- critically --- the span the query was generated from begins past token 512 in
30.7\% of rows and past token 1024 in 14.2\%.

\textbf{Decontamination.} Both \texttt{it-long\_doc} and MLDR-it derive from
Wikipedia. Using content-defined 13-word shingles (retaining shingles whose first
word hashes to $0 \bmod 16$, so selection depends on content rather than offset,
and flagging an article at $\geq 2$ shared shingles), we found 50{,}839 of 650{,}885
articles (\textbf{7.811\%}) overlapping the MLDR-it test corpus. These are excluded
from training and the exclusion list is released with the code.

\textbf{The gate.} Two phase-1-only models were trained on an identical
500{,}000-triplet mixture (250k mMARCO + 250k \texttt{it-long\_doc}), differing in
exactly one variable: document length, 512 versus 1024. The gate, fixed in advance,
required the 1024 arm to beat the 512 arm on the MLDR-it \emph{report} half by more
than the $0.0030$ noise floor, without regressing mMARCO-it or MIRACL-ita by more
than the same amount, and to win on best-of-arm as well as like-for-like.

\begin{table}[ht]
\centering
\small
\begin{tabular}{lccccl}
\toprule
\textbf{Comparison} & \textbf{512} & \textbf{1024} & \textbf{$\Delta$} & \textbf{$p$} & \textbf{Verdict} \\
\midrule
MLDR-it \ndcg{}, unchunked & 0.3895 & 0.4188 & $+0.0293$ & $0.35$ & not significant \\
MLDR-it \ndcg{}, chunked   & 0.4576 & 0.4544 & $-0.0032$ & $0.85$ & tied \\
mMARCO-it \mrr{} (guardrail) & --- & --- & $-0.0057$ & $0.12$ & not significant \\
MIRACL-ita \ndcg{} (guardrail) & --- & --- & $+0.0062$ & $0.46$ & not significant \\
\bottomrule
\end{tabular}
\caption{Training at 1024 tokens versus 512, identical data, report half
($n = 104$ MLDR-it queries). Neither primary comparison clears the noise floor with
significance, and best-of-arm favours the \emph{512} arm (0.4576 against 0.4544).
The gate fails.}
\label{tab:longdoc}
\end{table}

The informative part is not the gate result but the magnitudes around it: applying
chunking lifted the 512 arm by $+0.0681$ and the 1024 arm by $+0.0356$ --- both far
larger than the $\leq 0.03$ gap between the training arms. Training natively at
length added nothing once post-hoc chunking was available to both arms.

Taken with Section~\ref{sec:chunking}, two independent measurements now agree that
the long-document story here is an inference-time protocol question, not a
training-length question. We released the round-1 weights plus the chunking recipe
and abandoned this track.

\section{Efficiency}
\label{sec:efficiency}

Multi-vector indexes are the practical cost of late interaction, and chunking
multiplies it. Table~\ref{tab:efficiency} records what MLDR-it (10{,}000 documents,
200 queries) costs on one RTX 3090.

\begin{table}[ht]
\centering
\small
\begin{tabular}{lrrr}
\toprule
\textbf{Protocol} & \textbf{Vectors} & \textbf{Wall clock} & \textbf{Peak host RAM} \\
\midrule
Truncated 512, unchunked & $\approx$5.1M & 148\,s & $\approx$4\,GB \\
Chunked 2000/200, 4{,}000-char documents & $\approx$13.7M & 366\,s & 12\,GB \\
Chunked 2000/200, 12{,}000-char documents & $\approx$29.7M & 999\,s & 26\,GB \\
\bottomrule
\end{tabular}
\caption{Indexing and retrieval cost on MLDR-it. The bottom row sits at this
machine's 27\,GB ceiling for a 10{,}000-document corpus: the $+0.0602$ \ndcg{} from
chunking costs $6.7\times$ the wall clock and roughly $6\times$ the memory, and the
100{,}000-document mMARCO pool cannot be run chunked here at all.}
\label{tab:efficiency}
\end{table}

One infrastructure note with reproducibility consequences: chunking pushes the
document count past the harness's brute-force MaxSim threshold, which routes
retrieval into an approximate-nearest-neighbour path. On this machine that path was
silently broken (a native extension failing to load against the installed PyTorch,
falling back to a heavier index), and the fallback exhausted host memory and killed
both long-document runs before the flag was corrected. Approximate paths that
degrade silently are a real hazard when the only symptom is an out-of-memory kill.

\section{Limitations}
\label{sec:limitations}

\textbf{Single-seed training.} Every model here was trained once. Our $0.0030$
noise floor bounds endpoint jitter between adjacent checkpoints, not seed-to-seed
variance, which is certainly larger. Deltas in the 0.01--0.03 range on MLDR-it
should be read with that in mind; a second phase-1 run at a different seed
($\approx$6.5\,h) would be the cheapest way to close this.

\textbf{MLDR-it has 200 queries.} Our only clean out-of-domain benchmark is also
our smallest. This is why several comparisons in Table~\ref{tab:main} land as ties.

\textbf{Three of four suites are partially in-domain}, and two of those are
community machine translations rather than native Italian resources. Absolute
values on MIRACL-ita and SQuAD-ita should not be compared against published
numbers, and they favour our model relative to the baselines.

\textbf{Pooled mMARCO} inflates all absolute values on that suite; only the
ranking is meaningful.

\textbf{Protocols are not mixed safely.} The chunked MLDR-it number
(0.4610) is not comparable to the truncated numbers of the other systems in
Table~\ref{tab:main}, and we do not present it as such.

\textbf{Dataset revisions were not pinned} at training time (they are pinned for
the long-document source). Runs are reproducible on our machine from cache, not
necessarily elsewhere.

\textbf{Not attempted.} Rank fusion with BM25 is the most obvious cheap win and is
unbuilt: chunked ItColBERT and BM25 are statistically tied on MLDR-it while
disagreeing on individual queries, which is exactly the condition under which
fusion pays. A 64-dimension variant (halving index size) and an MTEB-style
community evaluation also remain open; there is currently no Italian MTEB.

\section{Conclusion}

We trained an Italian-specialised late-interaction retriever on one consumer GPU in
under 15 GPU-hours, and it is the strongest Italian-specialised multi-vector model
we could measure --- ahead of every general-purpose late-interaction baseline
except \texttt{mLateOn}, at between a half and a quarter of their parameter counts.
That result came almost entirely from two uncontroversial decisions: initialise
from a checkpoint that already retrieves, and distil from a single teacher while
selecting checkpoints on retrieval metrics rather than on the distillation loss.

The more transferable finding is the one we did not expect. Two rounds of training
aimed at the model's weakest benchmark --- self-mined hard negatives, then native
long-context training --- each failed a pre-registered gate, while a change that
required no training at all, chunking documents at query time, produced the largest
measured gain in the project. The deficit was in how documents were being read, not
in what the model had learned. On a constrained budget, measuring where the loss
actually comes from bought more than either additional training round, and we would
recommend that ordering to anyone building a language-specialised retriever under
similar constraints.

\section*{Availability}

Model weights: \url{https://huggingface.co/enricollen/ItColBERT}. Training code,
evaluation harness, per-query scores, decontamination list, and the full
experimental record including the rejected rounds:
\url{https://github.com/enricollen/it-colbert}. Released under Apache 2.0.


\begin{thebibliography}{99}

\bibitem[Bonifacio et al.(2021)]{bonifacio2021mmarco}
L.~H. Bonifacio, V.~Jeronymo, H.~Q. Abonizio, I.~Campiotti, M.~Fadaee, R.~Lotufo,
and R.~Nogueira.
\newblock mMARCO: A multilingual version of the MS MARCO passage ranking dataset.
\newblock \emph{arXiv preprint arXiv:2108.13897}, 2021.

\bibitem[Chaffin and Sourty(2025)]{chaffin2025pylate}
A.~Chaffin and R.~Sourty.
\newblock PyLate: Flexible training and retrieval for late interaction models.
\newblock In \emph{Proceedings of the 34th ACM International Conference on
Information and Knowledge Management (CIKM)}, pages 6334--6339, 2025.

\bibitem[Chen et al.(2024)]{chen2024bgem3}
J.~Chen, S.~Xiao, P.~Zhang, K.~Luo, D.~Lian, and Z.~Liu.
\newblock BGE M3-Embedding: Multi-lingual, multi-functionality, multi-granularity
text embeddings through self-knowledge distillation.
\newblock \emph{arXiv preprint arXiv:2402.03216}, 2024.

\bibitem[Clavi\'e(2024)]{clavie2024jacolbertv25}
B.~Clavi\'e.
\newblock JaColBERTv2.5: Optimising multi-vector retrievers to create
state-of-the-art Japanese retrievers with constrained resources.
\newblock \emph{arXiv preprint arXiv:2407.20750}, 2024.

\bibitem[Gao et al.(2021)]{gao2021gradcache}
L.~Gao, Y.~Zhang, J.~Han, and J.~Callan.
\newblock Scaling deep contrastive learning batch size under memory limited setup.
\newblock In \emph{Proceedings of the 6th Workshop on Representation Learning for
NLP}, 2021.

\bibitem[Jha et al.(2024)]{jha2024jinacolbertv2}
R.~Jha, B.~Wang, M.~G\"unther, S.~Sturua, M.~K. Akram, and H.~Xiao.
\newblock Jina-ColBERT-v2: A general-purpose multilingual late interaction
retriever.
\newblock In \emph{Proceedings of the 4th Workshop on Multilingual Representation
Learning (MRL)}, 2024.
\newblock arXiv:2408.16672.

\bibitem[Khattab and Zaharia(2020)]{khattab2020colbert}
O.~Khattab and M.~Zaharia.
\newblock ColBERT: Efficient and effective passage search via contextualized late
interaction over BERT.
\newblock In \emph{Proceedings of SIGIR}, 2020.
\newblock arXiv:2004.12832.

\bibitem[LightOn(2025)]{lighton2025colbertzero}
LightOn.
\newblock ColBERT-Zero: Training late interaction models from dense retrievers.
\newblock Technical report / blog post, 2025.
\newblock \url{https://huggingface.co/blog/lightonai/colbert-zero}.

\bibitem[Louis et al.(2024)]{louis2024colbertxm}
A.~Louis, V.~Saxena, G.~van Dijck, and G.~Spanakis.
\newblock ColBERT-XM: A modular multi-vector representation model for zero-shot
multilingual information retrieval.
\newblock \emph{arXiv preprint arXiv:2402.15059}, 2024.

\bibitem[Mixedbread(2025)]{mxbai2025edgecolbert}
Mixedbread AI.
\newblock mxbai-edge-colbert-v0: Small late interaction models for edge retrieval.
\newblock \emph{arXiv preprint arXiv:2510.14880}, 2025.

\bibitem[Reimers and Gurevych(2019)]{reimers2019sentencebert}
N.~Reimers and I.~Gurevych.
\newblock Sentence-BERT: Sentence embeddings using Siamese BERT-networks.
\newblock In \emph{Proceedings of EMNLP-IJCNLP}, 2019.
\newblock arXiv:1908.10084.

\bibitem[Santhanam et al.(2022a)]{santhanam2022colbertv2}
K.~Santhanam, O.~Khattab, J.~Saad-Falcon, C.~Potts, and M.~Zaharia.
\newblock ColBERTv2: Effective and efficient retrieval via lightweight late
interaction.
\newblock In \emph{Proceedings of NAACL}, 2022.
\newblock arXiv:2112.01488.

\bibitem[Santhanam et al.(2022b)]{santhanam2022plaid}
K.~Santhanam, O.~Khattab, C.~Potts, and M.~Zaharia.
\newblock PLAID: An efficient engine for late interaction retrieval.
\newblock \emph{arXiv preprint arXiv:2205.09707}, 2022.

\bibitem[Warner et al.(2024)]{warner2024modernbert}
B.~Warner, A.~Chaffin, B.~Clavi\'e, O.~Weller, O.~Hallstr\"om, S.~Taghadouini,
A.~Gallagher, R.~Biswas, F.~Ladhak, T.~Aarsen, N.~Cooper, G.~Adams, J.~Howard, and
I.~Poli.
\newblock Smarter, better, faster, longer: A modern bidirectional encoder for fast,
memory efficient, and long context finetuning and inference.
\newblock \emph{arXiv preprint arXiv:2412.13663}, 2024.

\bibitem[Zhang et al.(2023)]{zhang2023miracl}
X.~Zhang, N.~Thakur, O.~Ogundepo, E.~Kamalloo, D.~Alfonso-Hermelo, X.~Li, Q.~Liu,
M.~Rezagholizadeh, and J.~Lin.
\newblock MIRACL: A multilingual retrieval dataset covering 18 diverse languages.
\newblock \emph{Transactions of the Association for Computational Linguistics},
11:1114--1131, 2023.

\end{thebibliography}
\end{document}